\pdfoutput=1

\documentclass[fleqn,usenatbib,letters]{mnras}

\usepackage[T1]{fontenc}

\DeclareRobustCommand{\VAN}[3]{#2}
\let\VANthebibliography\thebibliography
\def\thebibliography{\DeclareRobustCommand{\VAN}[3]{##3}\VANthebibliography}

\usepackage{graphicx}   % Including figure files
\usepackage{amsmath}    % Advanced maths commands
\usepackage{amssymb}    % Extra maths symbols

\title[The unipolar solar flares]{
The unipolar solar flares as a manifestation of the `topological'
magnetic reconnection}

\author[Yu. V. Dumin \& B. V. Somov]{
Yurii V. Dumin,$^{1,2}$%
\thanks{E-mail: \href{mailto:dumin@pks.mpg.de}{dumin@pks.mpg.de} (YVD)}
and Boris V. Somov$^{1}$\thanks{Deceased}
\\
$^{1}$Sternberg Astronomical Institute of Lomonosov Moscow State University,
Universitetskii prosp. 13, Moscow, 119234 Russia\\
$^{2}$Space Research Institute of Russian Academy of Sciences,
Profsoyuznaya str. 84/32, Moscow, 117997 Russia}

\date{Accepted XXX. Received YYY; in original form ZZZ}

\pubyear{2023}

\begin{document}
\label{firstpage}
\pagerange{\pageref{firstpage}--\pageref{lastpage}}
\maketitle

\begin{abstract}
Solar flares~-- which are the most prominent manifestation of the solar
activity~-- typically manifest themselves as a single or a set of luminous
arcs (magnetic flux tubes) rooted in regions of opposite polarity in
the photosphere.
However, a careful analysis of the archival data by \textit{Hinode} satellite
sometimes reveals surprising cases of flaring arcs whose footpoints
belong to regions of the same polarity or to areas without any appreciable
magnetic field.
Despite the counterintuitive nature of this phenomenon, it can be reasonably
interpreted in the framework of the so-called `topological model' of magnetic
reconnection, where a magnetic null point is formed due to specific
superposition of influences from remote sources rather than by local current
systems.
As a result, the energy release propagates along a separator of the
flipping two-dome structure rather than along a fixed magnetic field line.
Therefore, the luminous arc needs not to be associated anymore immediately
with the magnetic sources.
Here, we report both observational cases of the above-mentioned type
as well as provide their theoretical model and numerical simulations.
% Not more than 200 words for Letters.
\end{abstract}


%%%%%%%%%%%%%%%%%%%%%%%%%%%%%%%%%%%%%%%%%%%%%%%%%%

%%%%%%%%%%%%%%%%% BODY OF PAPER %%%%%%%%%%%%%%%%%%

\section{Introduction}
\label{sec:Intro}

Solar flares represent the main form of the solar activity
\citep{Svestka_76,Sturrock_80,Priest_81,Priest_82,Tandberg_88,Phillips_91,%
Somov_13,Janvier_15}.
They usually manifest themselves as a set of the luminous arcs~-- associated
with the magnetic field lines~-- tracing a propagation of heat fluxes and
energetic particles from the sites of energy release to the lower layers of
atmosphere; e.g. a comprehensive review by \citet{Huang_18}.
Such arcs should be obviously rooted in the regions of opposite polarity
of the magnetic field, as usually confirmed by the corresponding magnetograms.

However, a careful analysis of the archival data on solar flares, outlined in
Section~\ref{sec:Observ}, enabled us to identify a few surprising cases when
the footpoints of the flaring arcs are located in the regions of the same
polarity or in the areas without any appreciable magnetic fields.
In other words, such arcs cannot be evidently associated with the magnetic
field lines.
Despite a counterintuitive nature of this phenomenon, as will be shown in
Section~\ref{sec:Theor}, it can be reasonably described by the so-called
`topological' mechanism of the magnetic reconnection, where the spot of
energy release is formed due to specific superposition of influences from
the remote magnetic sources, rather than by the local current systems.
As a result, the hot spot propagates approximately along a separator of
the global magnetic field configuration, which is not associated immediately
with the magnetic sources.

\section{Observational data}
\label{sec:Observ}

The primary observational material for our study was the archive by Solar
optical telescope (SOT) onboard \textit{Hinode} satellite
\citep{Kosugi_07,Tsuneta_08}, which is available in public domain.\footnote{%
\url{https://hinode.nao.ac.jp/en/for-researchers/qlmovies/top.html}}
Particularly, we analysed the patterns of emission in the chromospheric line
\ion{Ca}{ii}\,H, which is formed by a moderately heated plasma
($ {\approx}10^4 $~K, i.e. about twice the photospheric temperature).
We preferred to use this line because we were interested in the relatively
weak flares, where the magnetic field was not substantially disturbed by
the local plasma processes in the course of the flare development.
Besides, yet another advantage of the line \ion{Ca}{ii}\,H is that~--
along with the flaring arcs~-- it enables one to observe also the sunspots
and, thereby, to make some preliminary judgement about the structure of
magnetic field.

%%%%%%%%%%%%%%%%%%%%%%%%%%%%%%%%%%%%%%%%%%%%%%%%%%%%%%%%%%%%%%%%%%%%%%%%
\begin{figure*}
\includegraphics[width=0.85\textwidth]{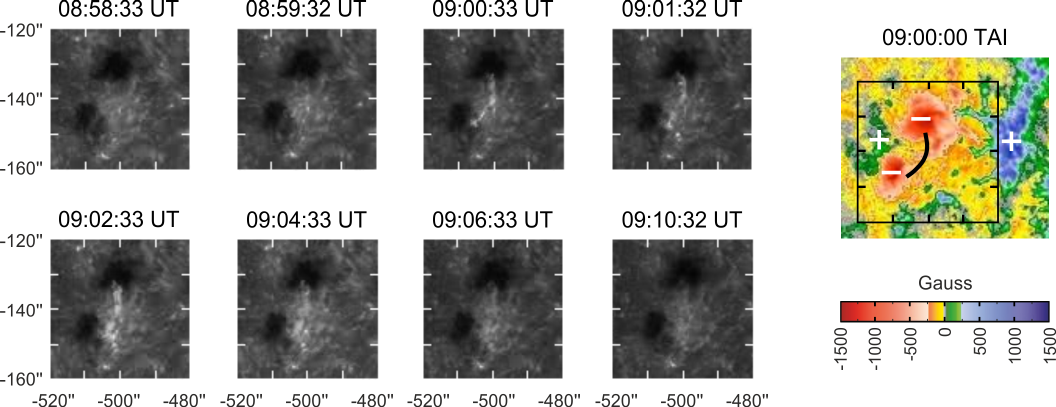}
\caption{Example of the unipolar microflare observed on 2014 October~1.
Left panels: Emission in \ion{Ca}{ii}\,H line recorded by \textit{Hinode}/SOT
at a few successive instants of time.
Right panel: Magnetogram of the corresponding region recorded by
\textit{SDO}/HMI (courtesy of NASA/SDO and the AIA, EVE and HMI science
teams).}
\label{fig:Observ_data}
\end{figure*}
%%%%%%%%%%%%%%%%%%%%%%%%%%%%%%%%%%%%%%%%%%%%%%%%%%%%%%%%%%%%%%%%%%%%%%%%

So, we performed initially a visual inspection of the large series of
images by \textit{Hinode}/SOT to find the cases of unusual location of
the flaring arcs with respect to the magnetic field sources.
Next, these case were analysed more carefully by comparing them with
the respective magnetograms recorded by HMI instrument onboard the
Solar dynamics observatory \citep[\textit{SDO},][]{Pesnell_12}, which are
also available online.\footnote{%
\url{http://jsoc.stanford.edu/HMI/hmiimage.html}}
The most interesting situations were observed in the regions of solar
surface involving complex geometrical configurations of the moderately-sized
sunspots with intermittent polarity.
Since they occurred most frequently in the periods of high solar activity,
we have analysed in most detail the \textit{Hinode}'s data for 2014 and 2015.

One of the interesting cases is presented in Fig.~\ref{fig:Observ_data}.
We can see here a small-scale ($ {\sim}15 $~arcsec) flaring arc, rooted
at the edges of two sunspots.
So, at the first sight, it looks like an ordinary magnetic field line.
Surprisingly, inspection of the corresponding magnetogram (right panel
in the same figure) shows that these sunspots are of the same (negative)
polarity.
Therefore, this arc cannot be evidently associated with the field line.

Although such unusual arcs are very seldom, they are identified rather
regularly under inspection of the sufficiently long time series of the images.
As will be shown in the next section, these phenomena can be well explained
by the so-called `topological' ignition of the magnetic reconnection.%
\footnote{To avoid misunderstanding, let us emphasize that the topological
methods are widely used in various contexts in the modern solar physics.
\citep[e.g., reviews][and references therein]{Longcope_05,Janvier_17}.
In the present paper, we shall call the `topological mechanism' only
the particular effect found by \citet{Gorbachev_88}.}

\section{Theoretical model}
\label{sec:Theor}

A commonly-accepted theory of the solar flares is based on the process of
magnetic reconnection, when the magnetic field lines break apart and then
reconnect again in a new configuration
\citep[e.g., monographs][and references therein]{Priest_00,Somov_13}.
This process develops in the null (or neutral) point of X-type, where all
components of the magnetic field vanish.
Such a null point is usually assumed to be formed by the local current
systems in the vicinity of the spot of reconnection.

%%%%%%%%%%%%%%%%%%%%%%%%%%%%%%%%%%%%%%%%%%%%%%%%%%%%%%%%%%%%%%%%%%%%%%%%
\begin{figure*}
\includegraphics[width=0.75\textwidth]{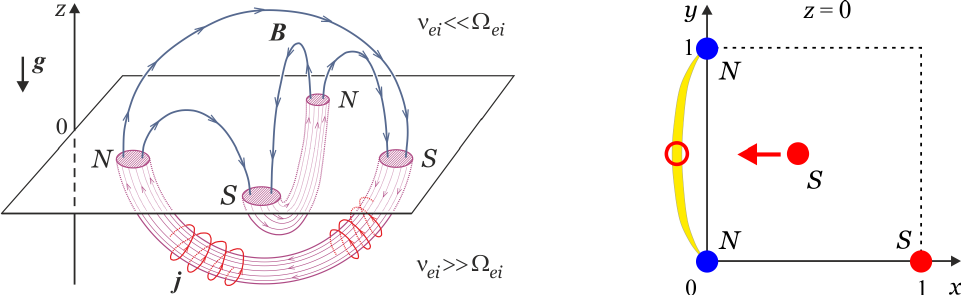}
\caption{Sketch of the `topological' model.
Left panel: formation of the effective `magnetic charges' at open ends of
the magnetic flux tubes.
Right panel: arrangement of the magnetic charges in $ z=0 $ plane at which
the `topological instability' occurs.}
\label{fig:GKSS_model}
\end{figure*}
%%%%%%%%%%%%%%%%%%%%%%%%%%%%%%%%%%%%%%%%%%%%%%%%%%%%%%%%%%%%%%%%%%%%%%%%

However, in principle, there might be also an absolutely different mechanism
for the formation of X-point, namely, due to specific superposition of
influences from the remote sources (sunspots), when the local currents are
absent at all.
The possibility of this option was pointed out for the first time over 30~years
ago by \citet{Gorbachev_88}, who employed some sophisticated theorems of
differential geometry and algebraic topology.
So, the corresponding mechanism was called the `topological trigger' of
magnetic reconnection.
Subsequently, this idea was developed in more detail by
\citet{Somov_08,Somov_13,OreshinaI_09,OreshinaA_12}.

As depicted in the left panel of Fig.~\ref{fig:GKSS_model}, the topological
model of reconnection is usually formulated in the approximation of potential
magnetic field formed by the point-like magnetic sources on (or slightly
below) the surface of photosphere
\citep[for more details, see][and references therein]{Longcope_96}.
From the physical point of view, such sources (effective magnetic charges)
represent open ends of the magnetic flux tubes, existing in the deeper
layers of the Sun, where characteristic collisional frequencies of both
electrons and ions~$ {\nu}_{ei} $ are much greater than their
gyrofrequencies~$ {\Omega}_{ei} $.
In the upper layers (above the photosphere), where the opposite inequality
$ {\nu}_{ei}\,{\ll}\,{\Omega}_{ei} $ holds, the electric
currents~$ \mathbfit{j} $ can no longer flow across the magnetic field.
So, in the first approximation, the field becomes potential and formed by
the effective magnetic charges (the open flux tubes):%
\footnote{
Of course, the potential approximation is inadequate for the large solar
flares, whose development in known to be associated with powerful local
electric currents; but it should work rather well for the microflares,
which are the main subject of the present study.}
\begin{equation}
\mathbfit{B} = -\nabla \varphi , \qquad
\varphi = \sum\limits_i \frac{e_i}{r_i} \, ,
\label{eg:Mag_field}
\end{equation}
where
$ e_i $~is the magnitude of the $i$'th magnetic charge, and
$ \mathbfit{r}_i $~is its radius vector.

%%%%%%%%%%%%%%%%%%%%%%%%%%%%%%%%%%%%%%%%%%%%%%%%%%%%%%%%%%%%%%%%%%%%%%%%
\begin{figure*}
\includegraphics[width=0.95\textwidth]{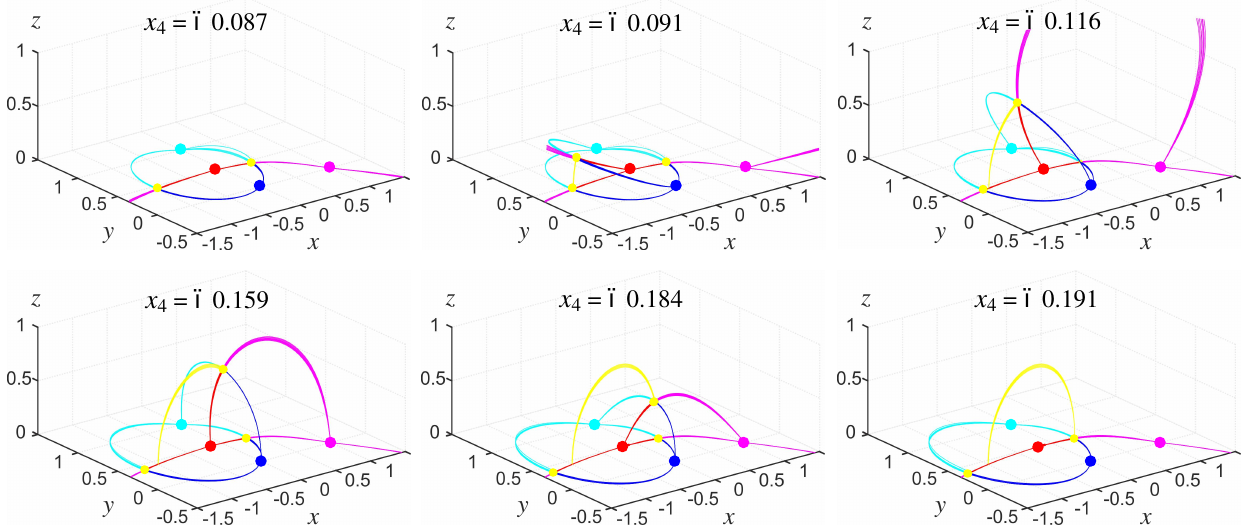}
\caption{Development of the `topological trigger', represented by
the flipping two-dome structure.
Red, magenta, blue and cyan balls are the magnetic sources;
and the field lines connecting them with the null points are marked by
the same colours.
Yellow balls are the null points of the magnetic field, and a yellow curve
is the trajectory of the bifurcated null point.
Coordinate~$ x_4 $ of the moving fourth magnetic charge (shown in red)
is indicated at the top of each panel.
The corresponding animated image is available as movie in the Supplementary
material.}
\label{fig:Flip_struct}
\end{figure*}
%%%%%%%%%%%%%%%%%%%%%%%%%%%%%%%%%%%%%%%%%%%%%%%%%%%%%%%%%%%%%%%%%%%%%%%%

In most cases, the pattern of the potential field lines is rather simple.
However, as was found by \citet{Gorbachev_88}, there are some specific
(`topologically unstable') configurations of the effective charges in
$ z = 0 $ plane where their tiny displacement results in a drastic change
of the entire magnetic field.
Moreover, such a reconfiguration is associated with birth and a fast motion of
the new null point of the field high above the plane of its sources.
Just this is a crucial prerequisite for the onset of magnetic reconnection.
(Let us emphasize once again that the above-mentioned null point is formed
without any local electric currents but solely due to the action by the remote
sources.)

In general, searching for the unstable topological configurations is a very
hard mathematical task, which was solved by now only for a few particular
situations.
In fact, the most studied case is the four-charge configuration of the slanted
\textit{T}-shape, shown in the right panel of Fig.~\ref{fig:GKSS_model}:
It is composed of two pairs of the opposite magnetic charges (two N(orth) and
two S(outh)) of the same magnitude.
Three of them are located at the vertices of a right equilateral triangle,
while the fourth charge moves in the same plane, i.e. can take an arbitrary
position.
For the sake of convenience, a cathetus of the triangle is chosen as the unit
of length, and absolute values of all magnetic charges~$ e_i $ are also
unitary.

Then, as was found by \citet{Gorbachev_88}, there is a narrow crescent region
of instability, schematically shown in yellow in the figure.
When the fourth charge enters this region (as shown by the open circle),
a global structure of the magnetic field sharply changes, and a new null point
should be born somewhere at $ z > 0 $.
Unfortunately, this fact has been proven in the above-cited work only as
an `existence theorem', and dynamics of the magnetic field in the course of
such reconstruction remained unclear for a long time.

In the present work, we performed a detailed numerical simulation of the
above-mentioned process, and its results are shown in
Fig.~\ref{fig:Flip_struct}.
Here, six panels illustrate evolution of the `topological skeleton' (i.e.,
a set of the field lines connecting the magnetic charges with null points)
in the course of development of the topological instability, when the forth
charge moves from the right to left along the line~$ (x_4, 0.5) $.
The corresponding values of its coordinate~$ x_4 $ are indicated at the top
of each panel.

The first panel ($ x_4 = -0.087 $) illustrates the magnetic field structure
just before the entrance into the zone of topological instability.
One can see here a small and a part of the large circle, which are the bases
of two domes separating four different types of topological connectivity
of the magnetic field lines \citep{Brown_99,Brown_01}.
There are two null points in the plane $ z = 0 $ (shown by the yellow balls),
which are of no interest for the magnetic reconnection because they are
localized in the dense photospheric plasmas.

The next four panels ($ x_4 = -0.091, -0.116, -0.159 $ and $ -0.184 $)
illustrate a development of the instability when the fourth source moves
through the yellow crescent region in Fig.~\ref{fig:GKSS_model}.
Then we can see in Fig.~\ref{fig:Flip_struct} that a new null point splits off
from the old one and begins to move quickly upward along the yellow arc.
As a result, the entire two-dome structure of the magnetic field experiences
a kind of flipping.
It is important to emphasize that all these transformations occur under a
tiny displacement of the fourth source, which is almost invisible in the scale
of the figure.
(The corresponding configuration of the magnetic charges, when the instability
occurs, looks like the slanted letter `\textit{T}' rotated by~$ \pi / 2 $.)

Finally, the last panel ($ x_4 = -0.191 $) corresponds to the end of the
instability, when the quickly moving null point merges with the second
previously-existing null point, and the magnetic skeleton restores its
unperturbed configuration.
The only noticeable change is a position of the first null point, which
is shifted to the left as compared to its original location.

%%%%%%%%%%%%%%%%%%%%%%%%%%%%%%%%%%%%%%%%%%%%%%%%%%%%%%%%%%%%%%%%%%%%%%%%
\begin{figure}
\includegraphics[width=0.85\columnwidth]{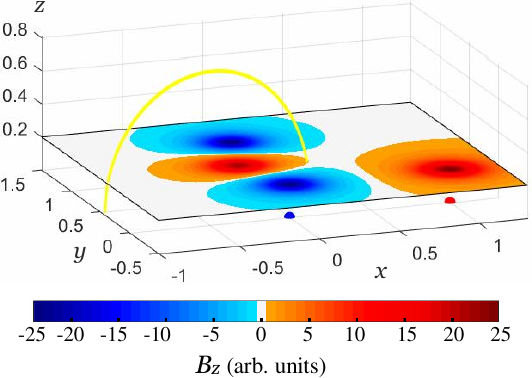}
\caption{Location of the simulated flaring arc with respect to the model
magnetogram (i.e., a vertical component of the magnetic field).
The observed photospheric surface is assumed to be located at height
$ z=0.2 $ above the plane of sources.}
\label{fig:Arc_vs_magnetogr}
\end{figure}
%%%%%%%%%%%%%%%%%%%%%%%%%%%%%%%%%%%%%%%%%%%%%%%%%%%%%%%%%%%%%%%%%%%%%%%%

Referring to the earlier studies on the topology of solar magnetic
fields, let us mention that our model exhibits approximately the same type
of bifurcation as the one presented in Fig.~9 of the paper by \citet{Brown_01},
where it was called a `local-double-separator bifurcation'.
Both before and after the stage of topological instability, there is
a separator at the intersection of two domes above the $ z = 0 $ plane, whose
feet are in the null points shown in yellow.
During the topological instability, there is an additional null
point, which splits off from the left null point and quickly moves to
the right one.
Thereby, the original separator splits into two segments.
To avoid misunderstanding, it should be emphasized that this situation is
qualitatively different from the bifurcations caused by the appearance or
disappearance of intersections between the separatrix surfaces (domes and
walls), e.g., as discussed by \citet{Brown_99}.

Does reconnection occur at a null point or at a separator?
The answer is both.
In our scenario the most favourable place for the reconnection and
the respective energy release is around the null point.
On the other hand, since the bifurcated null point moves quickly along
the separator, a luminous arc caused by the reconnection coincides
approximately with the separator.
(However, it should be kept in mind that~-- since the separator slightly
shifts with time~-- a trajectory of the null point does not coincide exactly
with the separator at any instant of time.
This fact is well seen in Fig.~\ref{fig:Flip_struct}, where a left footpoint
of the yellow arc slightly deviates from the left null point in the plane
$ z = 0 $.
We preferred to not show the separator in this figure to avoid confusion
with the null-point trajectory.)

From the observational point of view, it is important to emphasize
that trajectory of the bifurcated null point~-- along which the magnetic
reconnection and the energy release processes should develop~-- is absolutely
unrelated to the local configuration of the magnetic field lines.
This is illustrated in Fig.~\ref{fig:Arc_vs_magnetogr}, where the null-point
trajectory is superimposed onto the model magnetogram, namely, the vertical
component of the magnetic field~$ B_z $ in the plane $ z = 0.2 $.
(The observed photospheric surface was taken slightly above the plane of
sources for the sunspots to be of a finite size.)
As is seen here, one footpoint of the arc is located at the edge of a sunspot,
while another footpoint belongs to the area of very weak magnetic field.
Unfortunately, it is impossible to accurately compare this structure with
observational data in Fig.~\ref{fig:Observ_data} because of the large
uncertainty in the approximation of real magnetic field by the point-like
magnetic sources.

\section{Discussion and conclusions}
\label{sec:Discus}

In summary, simulations performed in the framework of topological model
in Section~\ref{sec:Theor} have shown that
\begin{enumerate}
\item
a trajectory of the bifurcated null point, where the energy release processes
are expected, is formed very quickly, and
\item
location of this trajectory is actually irrelevant to the configuration of
the local magnetic field lines.
\end{enumerate}

Both these features are in good qualitative agreement with the observational
data presented in Section~\ref{sec:Observ}:
Really, the `unipolar' flaring arc develops at the sufficiently short time
scale (less than 1~min, which is limited by temporal resolution of the
detectors) and extends between two sunspots of the same polarity, as was
already discussed in the Introduction.
Therefore, the model of `topological trigger' represents a reasonable
theoretical explanation for such kind of events.

It is rather surprising why the unipolar flares did not attract attention
before, while such events could be identified repeatedly under inspection of
the sufficiently long series of observations.
Most probably, they were disregarded \textit{a priori} as a kind of `noise',
since the flaring loops between the sources of the same polarity looked
unacceptable from the physical point of view.

Of course, an important question is: where does the flare energy come
from if we consider the potential field, i.e. the state of minimal energy
at the given boundary conditions?
One evident option is that the energy comes immediately from a redistribution
(shifts) of the sources.
Since plasmas in the upper layers are rarefied, the energy required to heat
them is relatively small as compared to the energy stored in the lower layers,
about $ z = 0 $.
So, a small shift of the sources (e.g., by 5--10 per cent) in the region of
topological instability will lead to a comparable change in the magnetic field
energy, and this may be sufficient to feed a microflare.
Yet another option might be that the real magnetic field is slightly
non-potential (force-free) but close to the topologies described here.
Really, as was shown by \citet{Brown_00}, topological structures of
potential and force-free fields with given boundary conditions are
usually very similar if the configuration of the sources is non-degenerate
(i.e., not too symmetric) and we do not consider very distant regions of
space.
Then, the force-free field will store excess energy, which could be
released in the reconnection.

Unfortunately, the theory of topological reconnection (especially,
identification of the unstable topological configurations) is a very hard
mathematical subject, which is sufficiently developed by now only for the case
of no more than four sources.
Bifurcations of the null points in some particular systems with multiple
sources were studied by \citet{Inverarity_99}; but they did not show, in
general, a fast motion of the null points.
Attempts to go beyond the point-charge and potential-field
approximations were undertaken by \citet{Longcope_01,Longcope_02}.
Furthermore, a promising approach to generalising the topological criteria
might be based on the Morse--Smale topological indices instead of
the Euler--Poincar{\'e} ones.
Such an idea was pursued in the recent work by \citet{Zhuzhoma_22}, but this
theory is only at the initial stage of development, and the results obtained
cannot be confronted yet with observations.

From the observational point of view, a considerable improvement in
the identification of unipolar flares can be achieved by utilising the
images in \ion{He}{i} 10\,830\,{\AA} line, e.g., taken by the New solar
telescope (NST) in the Big Bear Solar Observatory \citep{Zeng_16}.
This is a chromospheric line formed approximately at the same conditions as
\ion{Ca}{ii}\,H, but spatial resolution of NST is as high as~0.16\,arcsec,
which is substantially better than for \textit{Hinode}/SOT.
So, this might be a promising tool for the refined study of the unusual
microflares.
Unfortunately, there are no sufficiently long series of observations in this
line at the present time in the public domain.

\section*{Acknowledgements}

YVD is grateful to A.T.~Lukashenko and E.V.~Zhuzhoma for the discussion of
topological issues, to A.V.~Getling, A.V.~Oreshina and
I.~Slezak (former I.V.~Oreshina) for consultations on the processing of
magnetic fields, to V.F.~Vereshchagin for the advices on analysis of
\textit{Hinode} images,
and especially to the reviewer E.R.~Priest for many valuable suggestions.

\textit{Hinode} is a Japanese mission developed and launched by ISAS/JAXA,
with NAOJ as domestic partner and NASA and STFC (UK) as international partners.
It is operated by these agencies in co-operation with ESA and NSC (Norway).

Authors' contributions:
BVS suggested the general theoretical concept, supervised the work and
discussed the results.
YVD analysed the observational data, performed the corresponding numerical
simulations and prepared the manuscript.

%%%%%%%%%%%%%%%%%%%%%%%%%%%%%%%%%%%%%%%%%%%%%%%%%%

\section*{Data Availability}

The observational data analysed in the present paper, namely, the movies of
emission in \ion{Ca}{ii}\,H line recorded by \textit{Hinode}/SOT, are publicly
available in the Hinode QL Movie Archive; and the \textit{SDO}/HMI
magnetograms, at the web-site of the Joint Science Operations Center
(see footnotes in Section~\ref{sec:Observ}).
Computer software used for a simulation of the topological reconnection
can be obtained from YVD by reasonable request.


\bibliographystyle{mnras}
% \bibliography{Dumin}

\begin{thebibliography}{}
\makeatletter
\relax
\def\mn@urlcharsother{\let\do\@makeother \do\$\do\&\do\#\do\^\do\_\do\%\do\~}
\def\mn@doi{\begingroup\mn@urlcharsother \@ifnextchar [ {\mn@doi@}
  {\mn@doi@[]}}
\def\mn@doi@[#1]#2{\def\@tempa{#1}\ifx\@tempa\@empty \href
  {http://dx.doi.org/#2} {doi:#2}\else \href {http://dx.doi.org/#2} {#1}\fi
  \endgroup}
\def\mn@eprint#1#2{\mn@eprint@#1:#2::\@nil}
\def\mn@eprint@arXiv#1{\href {http://arxiv.org/abs/#1} {{\tt arXiv:#1}}}
\def\mn@eprint@dblp#1{\href {http://dblp.uni-trier.de/rec/bibtex/#1.xml}
  {dblp:#1}}
\def\mn@eprint@#1:#2:#3:#4\@nil{\def\@tempa {#1}\def\@tempb {#2}\def\@tempc
  {#3}\ifx \@tempc \@empty \let \@tempc \@tempb \let \@tempb \@tempa \fi \ifx
  \@tempb \@empty \def\@tempb {arXiv}\fi \@ifundefined
  {mn@eprint@\@tempb}{\@tempb:\@tempc}{\expandafter \expandafter \csname
  mn@eprint@\@tempb\endcsname \expandafter{\@tempc}}}

\bibitem[\protect\citeauthoryear{Brown \& Priest}{Brown \&
  Priest}{1999}]{Brown_99}
Brown D.,  Priest E.,  1999, \solphys, 190, 25

\bibitem[\protect\citeauthoryear{Brown \& Priest}{Brown \&
  Priest}{2000}]{Brown_00}
Brown D.,  Priest E.,  2000, \solphys, 194, 197

\bibitem[\protect\citeauthoryear{Brown \& Priest}{Brown \&
  Priest}{2001}]{Brown_01}
Brown D.,  Priest E.,  2001, \aap, 367, 339

\bibitem[\protect\citeauthoryear{Gorbachev, Kel'ner, Somov  \&
  Shvarts}{Gorbachev et~al.}{1988}]{Gorbachev_88}
Gorbachev V.,  Kel'ner S.,  Somov B.,   Shvarts A.,  1988, \sovast, 32, 308

\bibitem[\protect\citeauthoryear{Huang, Melnikov, Ji  \& Ning}{Huang
  et~al.}{2018}]{Huang_18}
Huang G.,  Melnikov V.,  Ji H.,   Ning Z.,  2018, Solar Flare Loops:
  Observations and Interpretations.
Springer, Singapore

\bibitem[\protect\citeauthoryear{Inverarity \& Priest}{Inverarity \&
  Priest}{1999}]{Inverarity_99}
Inverarity G.,  Priest E.,  1999, \solphys, 186, 99

\bibitem[\protect\citeauthoryear{Janvier}{Janvier}{2017}]{Janvier_17}
Janvier M.,  2017, J.\ Plas.\ Phys., 83, 535830101

\bibitem[\protect\citeauthoryear{Janvier, Aulanier  \& D{\'e}moulin}{Janvier
  et~al.}{2015}]{Janvier_15}
Janvier M.,  Aulanier G.,   D{\'e}moulin P.,  2015, \solphys, 290, 3425

\bibitem[\protect\citeauthoryear{Kosugi et~al.,}{Kosugi
  et~al.}{2007}]{Kosugi_07}
Kosugi T.,  et~al., 2007, \solphys, 243, 3

\bibitem[\protect\citeauthoryear{Longcope}{Longcope}{1996}]{Longcope_96}
Longcope D.,  1996, \solphys, 169, 91

\bibitem[\protect\citeauthoryear{Longcope}{Longcope}{2001}]{Longcope_01}
Longcope D.,  2001, Phys.\ Plas., 8, 5277

\bibitem[\protect\citeauthoryear{Longcope}{Longcope}{2005}]{Longcope_05}
Longcope D.,  2005, Liv.\ Rev.\ Sol.\ Phys., 2, 7

\bibitem[\protect\citeauthoryear{Longcope \& Klapper}{Longcope \&
  Klapper}{2002}]{Longcope_02}
Longcope D.,  Klapper I.,  2002, \apj, 579, 468

\bibitem[\protect\citeauthoryear{Oreshina \& Somov}{Oreshina \&
  Somov}{2009}]{OreshinaI_09}
Oreshina I.,  Somov B.,  2009, Astron.\ Lett., 35, 207

\bibitem[\protect\citeauthoryear{Oreshina, Oreshina  \& Somov}{Oreshina
  et~al.}{2012}]{OreshinaA_12}
Oreshina A.,  Oreshina I.,   Somov B.,  2012, \aap, 538, A138

\bibitem[\protect\citeauthoryear{Pesnell, Thompson  \& Chamberlin}{Pesnell
  et~al.}{2012}]{Pesnell_12}
Pesnell W.,  Thompson B.,   Chamberlin P.,  2012, \solphys, 275, 3

\bibitem[\protect\citeauthoryear{Phillips}{Phillips}{1991}]{Phillips_91}
Phillips K.,  1991, Vistas in Astronomy, 34, 353

\bibitem[\protect\citeauthoryear{Priest}{Priest}{1981}]{Priest_81}
Priest E.,  ed. 1981, Solar Flare Magnetohydrodynamics.
Gordon \& Breach, NY

\bibitem[\protect\citeauthoryear{Priest}{Priest}{1982}]{Priest_82}
Priest E.,  1982, Solar Magnetohydrodynamics.
D.~Reidel, Dordrecht

\bibitem[\protect\citeauthoryear{Priest \& Forbes}{Priest \&
  Forbes}{2000}]{Priest_00}
Priest E.,  Forbes T.,  2000, Magnetic Reconnection: MHD Theory and
  Applications.
Cambridge Univ. Press, Cambridge, UK

\bibitem[\protect\citeauthoryear{Somov}{Somov}{2008}]{Somov_08}
Somov B.,  2008, Astron.\ Lett., 34, 635

\bibitem[\protect\citeauthoryear{Somov}{Somov}{2013}]{Somov_13}
Somov B.,  2013, Plasma Astrophysics, Part II: Reconnection and Flares, 2 edn.
Springer, NY

\bibitem[\protect\citeauthoryear{Sturrock}{Sturrock}{1980}]{Sturrock_80}
Sturrock P.,  ed. 1980, Solar Flares: A Monograph from Skylab Solar
  Workshop~II.
Colorado Assoc. Univ. Press, Boulder

\bibitem[\protect\citeauthoryear{Svestka}{Svestka}{1976}]{Svestka_76}
Svestka Z.,  1976, Solar Flares.
D.~Reidel, Boston

\bibitem[\protect\citeauthoryear{Tandberg-Hanssen \& Emslie}{Tandberg-Hanssen
  \& Emslie}{1988}]{Tandberg_88}
Tandberg-Hanssen E.,  Emslie A.,  1988, The Physics of Solar Flares.
Cambridge Univ. Press, Cambridge, UK

\bibitem[\protect\citeauthoryear{Tsuneta et~al.,}{Tsuneta
  et~al.}{2008}]{Tsuneta_08}
Tsuneta S.,  et~al., 2008, \solphys, 249, 167

\bibitem[\protect\citeauthoryear{Zeng, Chen, Ji, Goode  \& Cao}{Zeng
  et~al.}{2016}]{Zeng_16}
Zeng Z.,  Chen B.,  Ji H.,  Goode P.,   Cao W.,  2016, \apjl, 819, L3

\bibitem[\protect\citeauthoryear{Zhuzhoma, Medvedev, Dumin  \& Somov}{Zhuzhoma
  et~al.}{2022}]{Zhuzhoma_22}
Zhuzhoma E.,  Medvedev V.,  Dumin Y.,   Somov B.,  2022, Physica D, 436, 133320

\makeatother
\end{thebibliography}

%%%%%%%%%%%%%%%%%%%%%%%%%%%%%%%%%%%%%%%%%%%%%%%%%%

% Don't change these lines
\bsp    % typesetting comment
\label{lastpage}

\end{document}